\documentclass[11pt]{article}
\usepackage{moriond,epsfig}

\def\apj{{\em ApJ}}
\def\mnras{{\em MNRAS}}

\def\be{\begin{equation}}
\def\ee{\end{equation}}
\def\bea{\begin{eqnarray}}
\def\eea{\end{eqnarray}}

\def\lesssim{\mathrel{\hbox{\rlap{\hbox{\lower4pt\hbox{$\sim$}}}\hbox{$<$}}}}
\def\gtrsim{\mathrel{\hbox{\rlap{\hbox{\lower4pt\hbox{$\sim$}}}\hbox{$>$}}}}

\begin{document}
\vspace*{4cm}
\title{MODELING GALAXY CLUSTERING\\ WITH COSMOLOGICAL SIMULATIONS}

\author{ANDREY V. KRAVTSOV}

\address{Department of Astronomy \& Astrophysics, The University of Chicago,\\
5640 S. Ellis Ave., Chicago, IL 60637, USA}

\maketitle

\abstracts{I review recent progress in understanding and modeling
  galaxy clustering in cosmological simulations, with emphasis on models
  based on high-resolution dissipationless simulations. During the
  last decade, significant advances in our understanding of abundance
  and clustering of dark matter halos allowed construction of
  accurate, quantitative models of galaxy clustering both in linear
  and non-linear regimes. Results of several recent studies show that
  dissipationless simulations with a simple, non-parametric model for
  the relation between halo circular velocity and luminosity of the
  galaxy they host, ${V_{\rm max}-L}$, predict the shape, amplitude,
  and luminosity dependence of the two-point correlation function in
  excellent agreement with the observed galaxy clustering in the SDSS
  data at $z\sim 0$ and in the DEEP2 samples at $z\sim 1$ over the
  entire probed range of projected separations.  In particular, the
  small-scale upturn of the correlation function from the power-law
  form in the SDSS and DEEP2 luminosity-selected samples is reproduced
  very well. At $z\sim 3-5$, predictions also match the observed shape
  and amplitude of the angular two-point correlation function of
  Lyman-break galaxies (LBGs) on both large and small scales,
  including the theoretically predicted strong upturn at small scales.
  This suggests that, like galaxies in lower redshift samples, the
  LBGs are fair tracers of the overall halo population and that their
  luminosity is tightly correlated with the circular velocity (and
  hence mass) of their dark matter halos.  }

\section{Introduction}
\label{sec:intro}

During the last decade, large observational surveys of galaxies both
at low and high redshifts have tremendously improved our knowledge of
galaxy clustering, its evolution, and the relation between the galaxy
and matter distributions.  A coherent picture has emerged in which
bright galaxies are strongly biased with respect to the matter
distribution at high redshifts,\cite{Steidel98,Giavalisco98,Adelberger03,Adelberger05,Ouchi04b,Ouchi05,Lee05,Hamana05} and in which the bias decreases with time in such a way that the amplitude of
galaxy clustering is only weakly evolving,\cite{Ouchi04b} as expected
in hierarchical structure formation.\cite{colin_etal99,kauffmann_etal99} The bias is also in
general scale-, luminosity-, and color-dependent.  Bright (red)
galaxies are more strongly clustered than faint (blue) galaxies both
in the local universe \cite{Norberg02a,Zehavi04,Zehavi05} and in the
distant past.\cite{Coil04,Coil05,pollo_etal06}

These trends are in general consistent with the picture in which
galaxies reside in extended dark matter (DM) halos, forming via
hierarchical collapse and merging of peaks in the initial density
field. Thus, for example, the stronger clustering of brighter galaxies
can be readily understood if they tend to populate more massive halos,
which are expected to be more clustered.\cite{kaiser84,mo_white96,sheth_tormen99,warren_seljak04}  Like
galaxies, the halos are strongly clustered at high redshifts and their
clustering strength evolves only weakly with time.\cite{colin_etal99,kauffmann_etal99}

Clustering of halos
of a given mass, formed in the standard structure formation scenario
from a gaussian initial density field, is simple at large, linear
scales, where it can be described by a single number --- the linear
bias.\cite{scherrer_weinberg99} Recently, it has been shown that in
addition to the mass dependence, the halo bias depends on other halo
properties, such as its formation time and mass
concentration.\cite{gao_etal05,harker_etal06,wechsler_etal06}
At small scales, the clustering of halos is more complicated and
bias is scale-dependent,\cite{colin_etal99,kravtsov_klypin99} the behavior
resulting from dynamical evolution of halos in high-density 
environments.\cite{kravtsov_klypin99,zentner_etal05} The non-linear clustering
of halos on small scales and the processes shaping it 
have been extensively investigated during the last decade
using dissipationless simulations.\cite{brainerd_villumsen92,brainerd_villumsen94,kravtsov_klypin99,kravtsov_etal04,Neyrinck04,Neyrinck05,conroy_etal06}

One of the most important theoretical advances of the last several
years is development of the Halo Model (HM) framework, in which galaxy
clustering on both linear and nonlinear scales is described
quantitatively using spatial and mass distribution of DM halos
calibrated against cosmological simulations and the Halo Occupation
Distribution (HOD) --- the probability distribution for a halo of mass
$M$ to host $N$ galaxies with specified properties (e.g., luminosity,
color, etc.). In its simplest form, the HOD is assumed to depend
solely on the halo mass. This appears to be a fairly good approximation.\cite{zentner_etal05}
However, in general the HOD can depend on other
halo properties or large-scale halo environment.

This framework proved to be extremely useful both in theoretical 
forecasts and interpretation of observed clustering data. For example, 
in the halo model the two-point correlation function is a sum of two separate
contributions: the one-halo term, which arises from pairs of galaxies
within the same dark matter halo, and the two-halo term, which
arises from pairs of galaxies from two different halos.
\cite{berlind_weinberg02,Cooray02} The one-halo contribution dominates on
small scales, while at scales larger than the size of the largest
virialized regions clustering is due to the two-halo term. In general, the two
terms are not  expected to combine so as to give a perfect power-law
correlation function. Departures from power-law are thus generically
expected in this model. In addition, the model {\em predicted}
that deviations of the correlation function from a
power-law should be even stronger at higher redshifts.\cite{Zheng04,kravtsov_etal04} 
This is because at high redshifts the merger rate is higher
and halos are more likely to have massive subhalos of comparable
mass and luminosities. When the merger rate decreases, such massive
companions disappear as they merge due to dynamical friction. 
These predictions have now been convincingly confirmed
by observations both at $z=0$ and higher redshifts.\cite{Zehavi04,Zehavi05,Coil05,Adelberger05,Ouchi05,Lee05,Hamana05}

Although the general idea of a galaxy-halo connection is definitely
reasonable and is not disputed, the key question is how tight this
connection is and whether properties of halos are tightly related to
the properties of galaxies they host.  Theoretical models of galaxy
clustering, partially reviewed here, and their comparison to the
wealth of  current observational data start to shed light on these
fundamental questions of galaxy formation theory.

\section{Theoretical Models of Galaxy Clustering}
\label{sec:models}

Ultimately, one would like to simulate the distribution of galaxies in a
large, representative volume of the Universe, while reliably and
self-consistently modeling their internal properties at the same time.
Given that this is not yet feasible with the current state of our
understanding of galaxy formation and the capabilities of the most
powerful supercomputers, some phenomenological modeling and
assumptions have to be made.  Historically, galaxy clustering in
cosmological simulations has been modeled using a variety of approaches.

The most direct approach is to use cosmological simulations which
include both dark matter and baryonic components, as well as galaxy
formation processes of radiative dissipation and phenomenological
recipes for star formation and stellar
feedback.\cite{pearce_etal99,katz_etal99,blanton_etal99,berlind_etal03,zheng_etal05}
Although properties of galaxies are not yet modeled reliably in such
simulations, they allow for unambiguous identification of galaxies as
dense clumps of gas and stars, as well as measurement of basic galaxy
properties such as stellar and baryonic mass, stellar ages,
luminosities and colors. These observables, in turn, allow for
extensive comparisons with observations. The main disadvantange is
that such simulations are generally computationally expensive, forcing
one to sacrifice the size of the simulated volume or spatial
resolution to make the simulations feasible.

The most popular approach to modeling galaxy clustering employs
semi-analytic modeling\cite{White91,Kauffmann94,Cole94,AvilaReese98,Somerville99,Cole00,Croton05,Bower05}, which uses phenomenological
recipes for specifying when, where, and how galaxies form and evolve
within dark matter halos, in conjunction with high-resolution
dissipationless simulations modeling spatial distribution and merger
histories of dark matter halos. Such hybrid models of increasing
degree of sophistications have been used to model evolution of galaxy
clustering, as well as trends with luminosity, color, and other galaxy
properties.\cite{kauffmann_etal99,benson_etal00,berlind_etal03,springel_etal05,Croton05}
These methods provide flexibility to explore the dependence of predictions
on particular assumptions about galaxy formation physics, albeit at
the expense of a fairly large number of free parameters, assumptions,
and (often uncertain) parameterizations of the complex physical
processes.

\begin{figure}
\vspace{-0.5cm}
\begin{center}
\psfig{figure=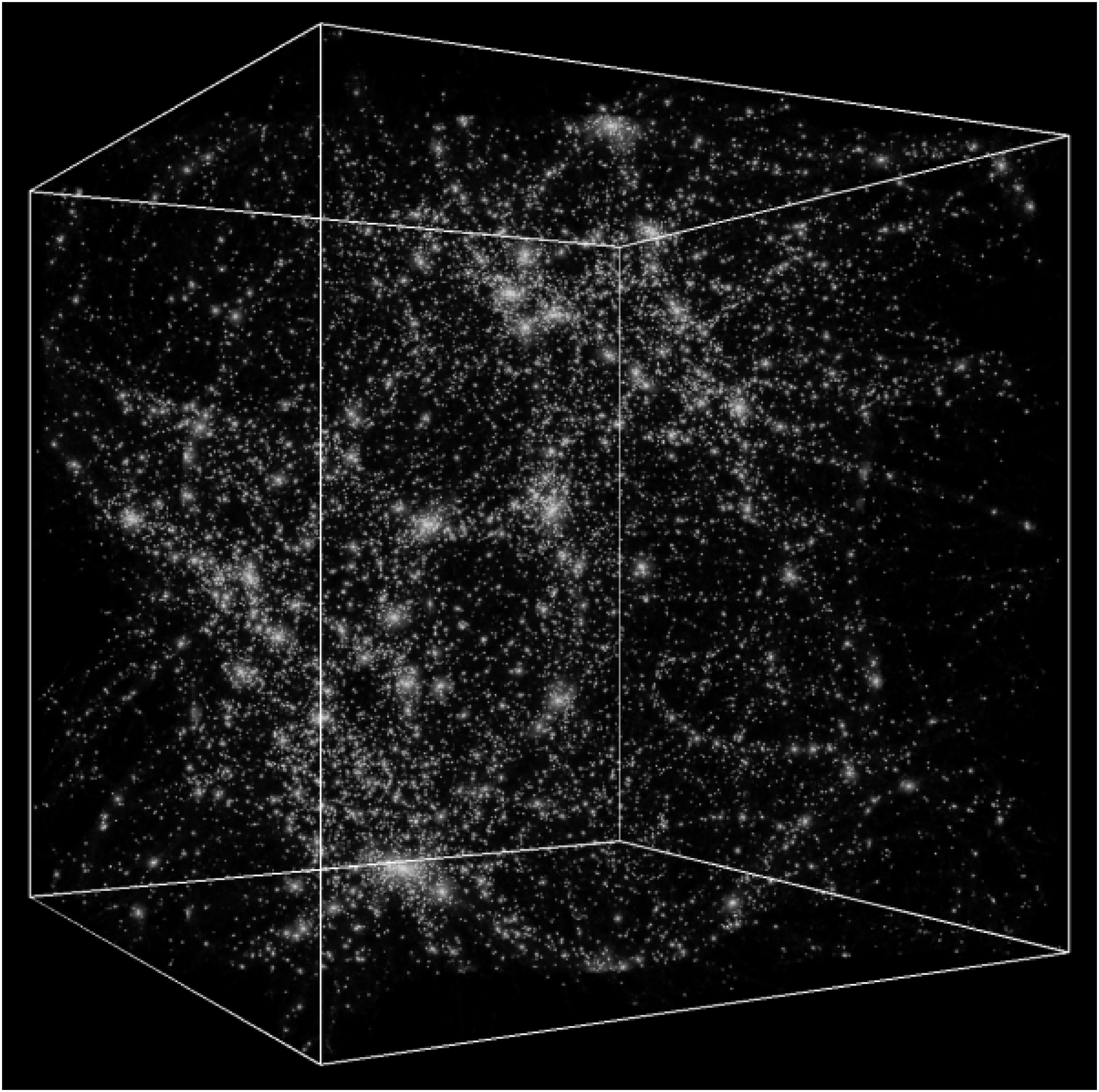,height=3in}
\psfig{figure=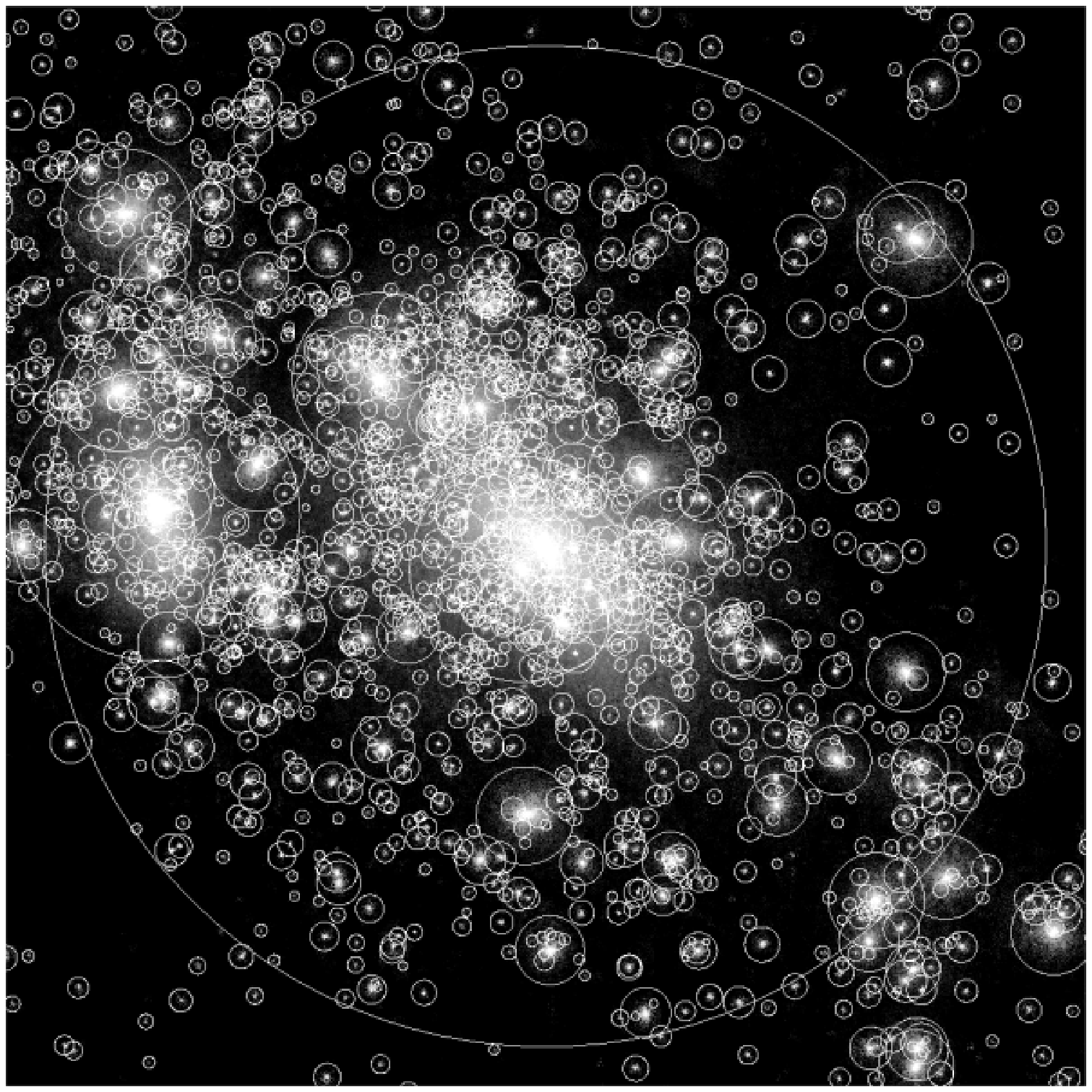,height=3in}
\end{center}
\caption{Left: Dark matter distribution in a flat $\Lambda$CDM simulation of a $60h^{-1}$~Mpc volume. Particles are color-coded on a greyscale according to the logarithm of the local density. 
One can see a network of filaments interconnecting groups and clusters. The network is filled
with small, dense dark matter halos. Right: distribution of dark matter within the virial
radius of a cluster-sized halo. The virial radius of the cluster is shown by the large
circle. The volume enclosed by the cluster virial radius is filled with smaller dense
{\em subhalos}, which are bound to the cluster and orbit within its potential. The circles 
indicate the individual objects identified by an automated halo finder (see Kravtsov et al. 2004 for details); the radii of the circles are proportional to the subhalo maximum circular
velocity ($r_h\propto V_{\rm max}\propto M_h^{1/3}$). 
\label{fig:dmsub}}
\end{figure}

The third, considerably simpler approach, is to use high-resolution 
dissipationless, DM-only simulations capable of following the evolution of both 
isolated, distinct halos and subhalos --- the bound, self-gravitating dark
matter clumps orbiting in the potential of their host halo.\cite{colin_etal99,kravtsov_klypin99,Neyrinck04,kravtsov_etal04,conroy_etal06} 
Subhalos are the descendants of halos accreted by a given system
throughout its evolution, which retain their identity in the face of
disruption processes such as tidal heating and dynamical
friction (see Figure~\ref{fig:dmsub}). In the context of
galaxy formation, there is little conceptual difference between halos
and subhalos, because the latter have also been genuine halos and
sites of galaxy formation in the past, before their accretion onto a
larger halo. We thus expect that each subhalo of sufficiently large
mass should host a luminous galaxy and this is indeed supported by
self-consistent cosmological simulations.\cite{Nagai05}
The observational counterparts of subhalos are then galaxies in
clusters and groups or the satellites around individual galaxies.  In
this sense, I will use the term {\it halos} to refer to both distinct
halos (i.e., halos not located within the virial radius of a larger
system) and subhalos.

With the assumptions that 1) there is one-to-one correspondence between
halos in dissipationless simulations and luminous galaxies and 2)
there is a tight relation between halo properties and properties of
the galaxy they host, one can use halo distribution self-consistently
modeled in such simulations to make detailed predictions for galaxy
clustering. Confronting these predictions with observations can test
the validity of the assumptions above and constrain the relations
between galaxies and their halos. In the remainder of this contribution, I will
describe such studies in more detail. Their results show that this simple
approach is remarkably accurate in describing the luminosity-dependence
and evolution of galaxy clustering.

\section{Modeling galaxy clustering in dissipationless simulations}

\subsection{Relating galaxies and halos}
\label{sec:ghrel}

Assuming that all luminous galaxies live in the dark matter halos 
identified in high-resolution dissipationless simulations, the key
question is how we relate a galaxy of a given luminosity to a specific
halo.  In what follows, I will consider only luminosity among the
possible properties of galaxies. Although it may be possible to relate
other properties, such as color, to halo properties, these relations are considerably more
uncertain and are likely to exhibit large scatter. Luminosities
(especially in the red and infrared bands), on the other hand, are
related to the total stellar mass of the galaxies.  The stellar
masses, in turn, can be expected to be related to the depth of the
potential well of the halo and hence to its maximum circular velocity
$V_{\rm max}$ (note that potential energy of a halo scales as
$W\propto V_{\rm max}^2$). This relation is thought to be the basis
of the observed Tully-Fisher relation. 
Indeed, tight relations between halo circular velocity and stellar 
mass of the galaxies they host are expected in all models of galaxy
formation. 

\begin{figure}
\psfig{figure=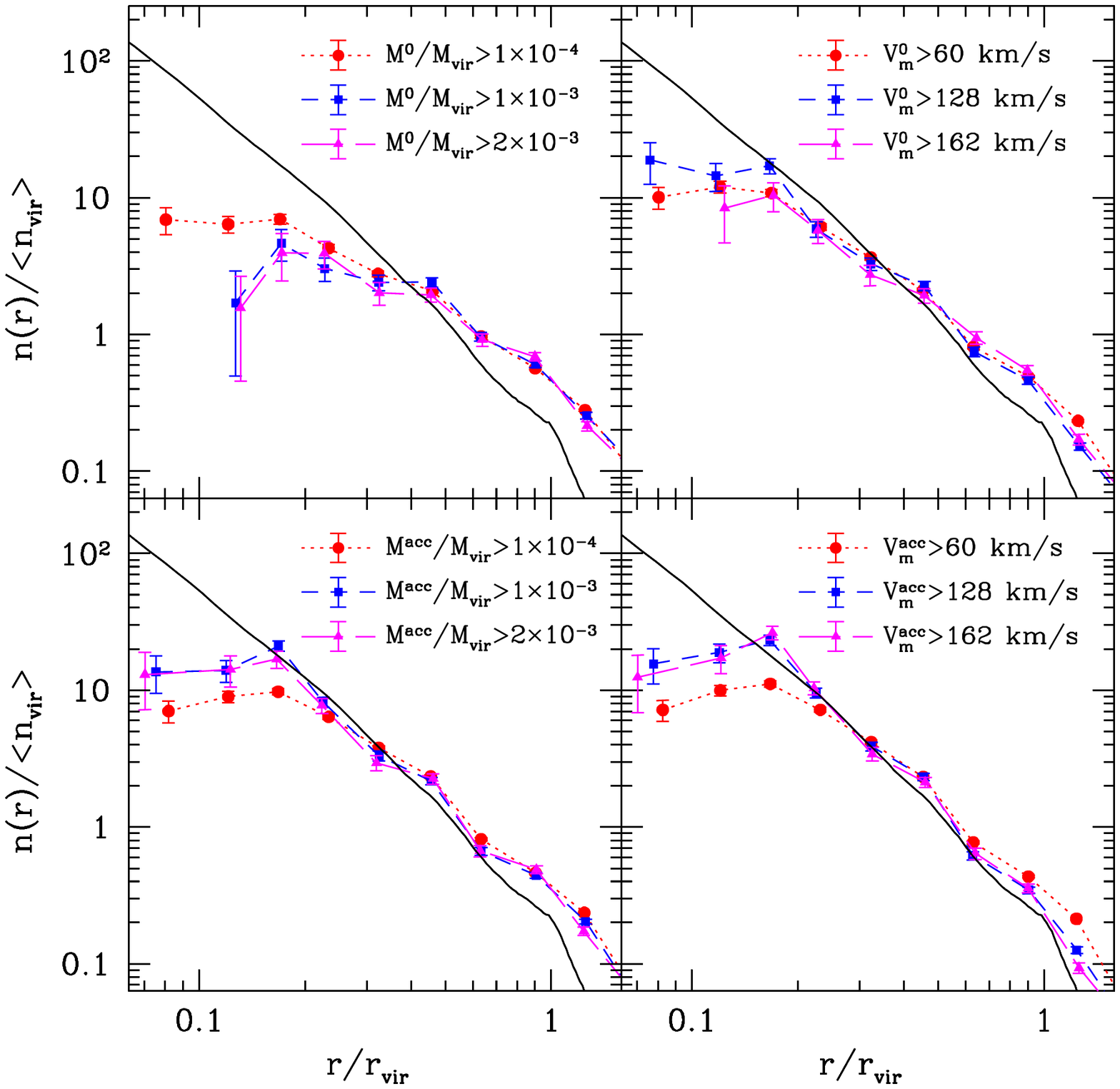,height=3in}
\psfig{figure=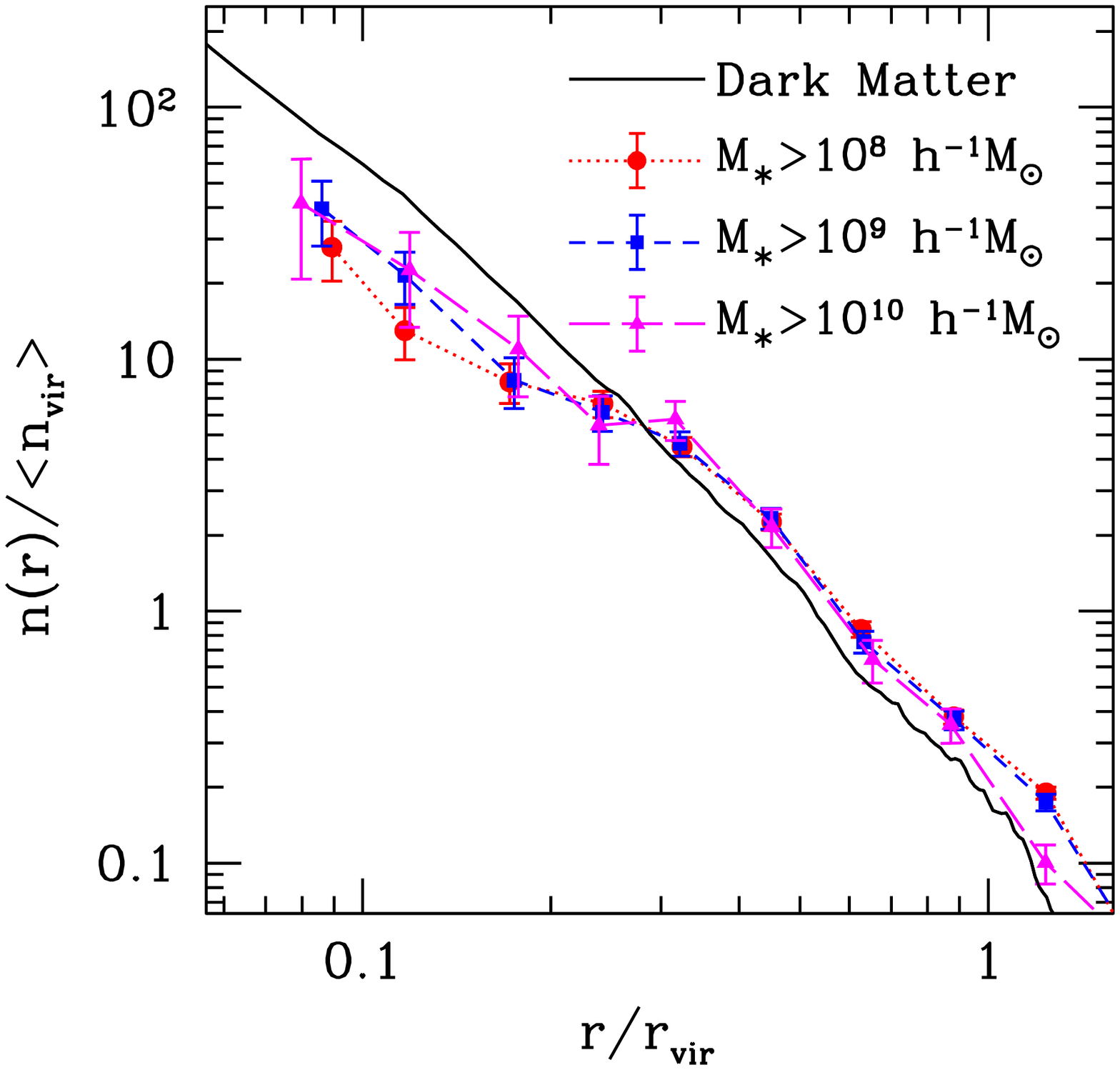,height=3in}
\vspace{-0.5cm}
\caption{Left: radial distribution of subhalos in high-resolution dissipationless
  simulations of cluster-sized halo. The solid line shows the density
  profile of dark matter. Different panels show radial number density
  profiles of subhalo samples selected using different criteria and
  different mass and circular velocity thresholds. In the top panels
  subhalos are selected using total bound mass and circular velocity
  at the present epoch, while in the bottom panels the mass and
  circular velocity values at the epoch at which each subhalo was
  accreted by the cluster are used. Right: radial distribution of
  subhalos in $N$-body$+$ hydro simulation of the same cluster with
  cooling and starformation. In this case, subhalos are selected using
  stellar mass of the galaxy they host. Comparison of the radial
  distribution in the left and right plots shows that at radii
  $r/r_{\rm vir}\gtrsim 0.15-0.20$ the selection using circular
  velocity at accretion mimicks selection using galaxy stellar
  mass.\hspace{9cm} Reproduced from Nagai \& Kravtsov (2005).
\label{fig:nk05}}
\end{figure}

The use of $V_{\rm max}$ as a halo property in simulations is
attractive because it is unambiguous both for distinct halos and
subhalos, which is not the case for the total mass.  It should be
noted that $V_{\rm max}$ measured in dissipationless simulations will
not correspond directly to observed rotation velocity of galaxies
because dissipationless simulations do not take
into account the effect of baryon condensation on circular
velocity.\cite{Blumenthal86,gnedin_etal04} For our purposes, however, 
it is sufficient that a monotonic correlation between luminosity 
and halo $V_{\rm max}$ is expected. 

\begin{figure}
\vspace{-0.5cm}
\begin{center}
\psfig{figure=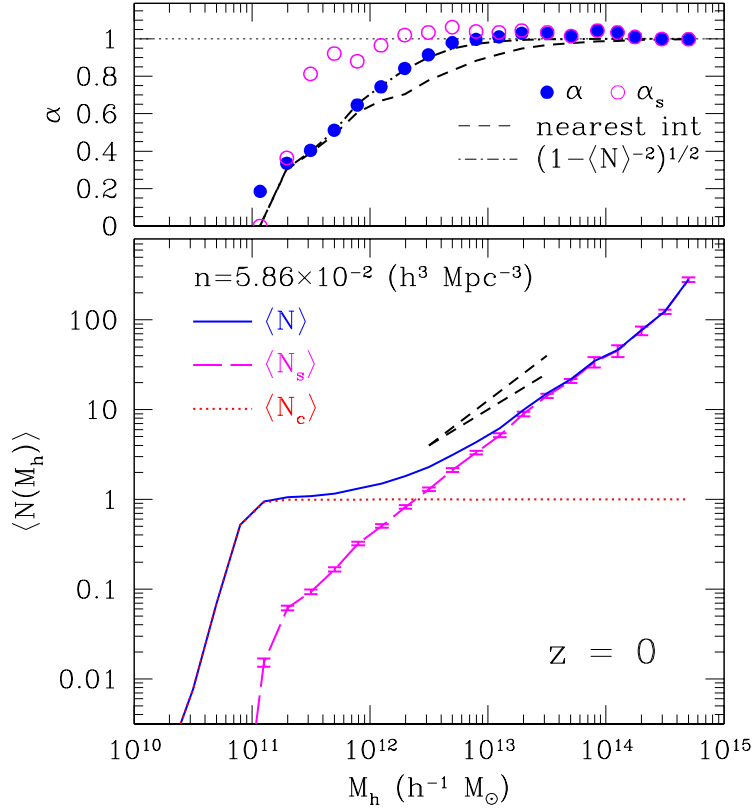,height=4.5in}
\end{center}
\vspace{-0.5cm}
\caption{Bottom panel: The subhalo occupation distribution --- the
mean number of subhalos with maximum circular velocities $V_{\rm max}>70$~km/s 
as a function of host halo mass. The solid line
shows the mean total number of halos including the hosts (i.e., for
each halo $N$ is the number of subhalos plus one, the host halo
itself), while the long-dashed line shows the mean number of
satellite halos.  Upper panel: the ratio of the square root of the
second HOD moment to the first moment, $\alpha\equiv \langle
N(N-1)\rangle^{1/2}/\langle N\rangle$, for the full HOD (solid
points) and the HOD of satellite halos (open points). The dotted
line at $\alpha=1$ shows the case of the Poisson distribution. Note
that the HOD becomes sub-Poisson at small host masses. However, the
HOD of satellites remains close to Poisson down to masses an order
of magnitude smaller than for the full HOD. The dot-dashed line
shows prediction for $\alpha$ of the total HOD, if the halo
occupation distribution of satellite subhalos is described by the
 Poisson distribution.  Reproduced from Kravtsov et al. (2004).
\label{fig:hod}}
\end{figure}

The existence of such a monotonic relation allows for a simple,
non-parametric model relating halo $V_{\rm max}$ and galaxy
luminosity. Specifically, the $V_{\rm max}-L$ relation is derived from
the measured abundance of halos as a function of their circular
velocity and the requirement that the relation matches the observed
luminosity function of galaxies. In the simplest version of the model,
no scatter is assumed and the $V_{\rm max}-L$ relation is derived by
matching the circular velocity and luminosity functions: $n(>V_{\rm max})=n(>L)$.

Such a simple model is bound to be too simplistic. First, there is certainly
scatter between observed galaxy luminosity and their halo circular
velocities.  Such scatter can be added into the model at the expense
of introducing a free parameter.\cite{tasitsiomi_etal05}
It is interesting that the amount of scatter between luminosity and
$V_{\rm max}$ may be constrained by joint comparisons of the model
predictions for galaxy-galaxy and galaxy-mass correlations because the
two statistics have different sensitivity to scatter and this
sensitivy depends on luminosity.\cite{conroy_etal06}

Second, circular velocity of halos measured at the epoch of
observations is not expected to be the optimal choice for subhalos.
For distinct isolated halos, the current circular velocity is a
measure of their potential well assembled during evolution, and can
therefore be expected to be tightly correlated with the stellar mass
(or more generally the baryonic mass) of the galaxy the halo hosts.
The circular velocity of subhalos in dissipationless simulations, on
the other hand, is a product of both mass buildup during the period
when the halo evolved in isolation {\it and} tidal mass loss, with an
associated decrease of $V_{\rm max}$, after the halo starts to orbit
within the virialized region of a larger object and experience strong
tidal forces.\cite{Hayashi03,Kravtsov04b,Kazantzidis04b} The stellar
component of galaxies in centers of halos, which should be more
tightly bound than halo dark matter, should be less affected by
tidal forces and can stabilize the mass distribution (and hence
$V_{\rm max}$) in the inner regions. We can therefore expect that
luminosity and stellar mass of galaxies hosted by halos in
dissipationless simulations should be correlated with the subhalo mass
or circular velocity, $V_{\rm max}^{\rm acc}$, {\it at the epoch of
  accretion}, rather than with its current value.

This is borne out by cosmological simulations that include gas
dynamics, cooling, and star formation.\cite{Nagai05} Such simulations
show that selection using $V_{\rm max}^{\rm acc}$ results in subhalo
distribution similar to the selection based on stellar mass of
galaxies hosted by subhalos (see Figure~\ref{fig:nk05}).  One can therefore
argue that a reasonable approach is to relate galaxy luminosity to the
current halo circular velocity for distinct halos and to the circular
velocity at accretion for subhalos.\footnote{The distinction between
  circular velocity at accretion and current epoch has little effect
  at $z\gtrsim 1$. This is because both the accretion and disruption
rates are high at high redshifts. The accreted halos do not survive
for a prolonged period of time, so that at each high-$z$ epoch most
of identified subhalos are recently accreted objects, which are yet
to experience significant tidal mass loss.}  The models based 
on the halo properties at accretion were recently used in several 
studies.\cite{Nagai05,vale_ostriker06,conroy_etal06,wang_etal06,berrier_etal06}
The results discussed
below show that such simple luminosity assignment model reproduces the
luminosity-dependence of galaxy clustering at different epochs with
remarkable, and perhaps surprising, accuracy.

\subsection{(Sub)Halo Occupation Distribution}
\label{sec:hod}

As I noted in \S~\ref{sec:intro}, the halo occupation distribution
plays a central role in the modeling of galaxy clustering in the
framework of the halo model. Given the assumption that every galaxy
corresponds to a DM halo in high-resolution simulation, it is
interesting to ask what form of the HOD the simulations predict.
Analysis shows that for samples of halos with circular velocities
larger than a certain threshold, the HOD is quite simple and can be
parameterized with only a few
parameters.\cite{kravtsov_etal04,zheng_etal05}

Specifically, the HOD can be understood as a combination of the
probability for a halo of mass $M$ to host a central galaxy,
associated with the halo itself and the probability to host a given
number $N_s$ of satellite galaxies, associated with the subhalos
(Figure~\ref{fig:hod}). Such logical division makes physical sense, 
because central galaxies occupy a special location in the halo near
the minimum of the potential well. Observational analogs of the
central galaxies are, for example, the Milky Way with its system of
satellites or a cD galaxy in a galaxy cluster. The HOD of the central
galaxies can be approximated by a step-like function, while the
satellite HOD can be well approximated by a Poisson distribution,
fully specified by its first moment. The first moment of the satellite
HOD can be well described by a simple power-law $\langle N_s\rangle
\propto M^{\beta}$ with $\beta\approx 1$ for a wide range of number
densities, redshifts, and different power spectrum normalizations.

\begin{figure}
\vspace{-0.5cm}
\psfig{figure=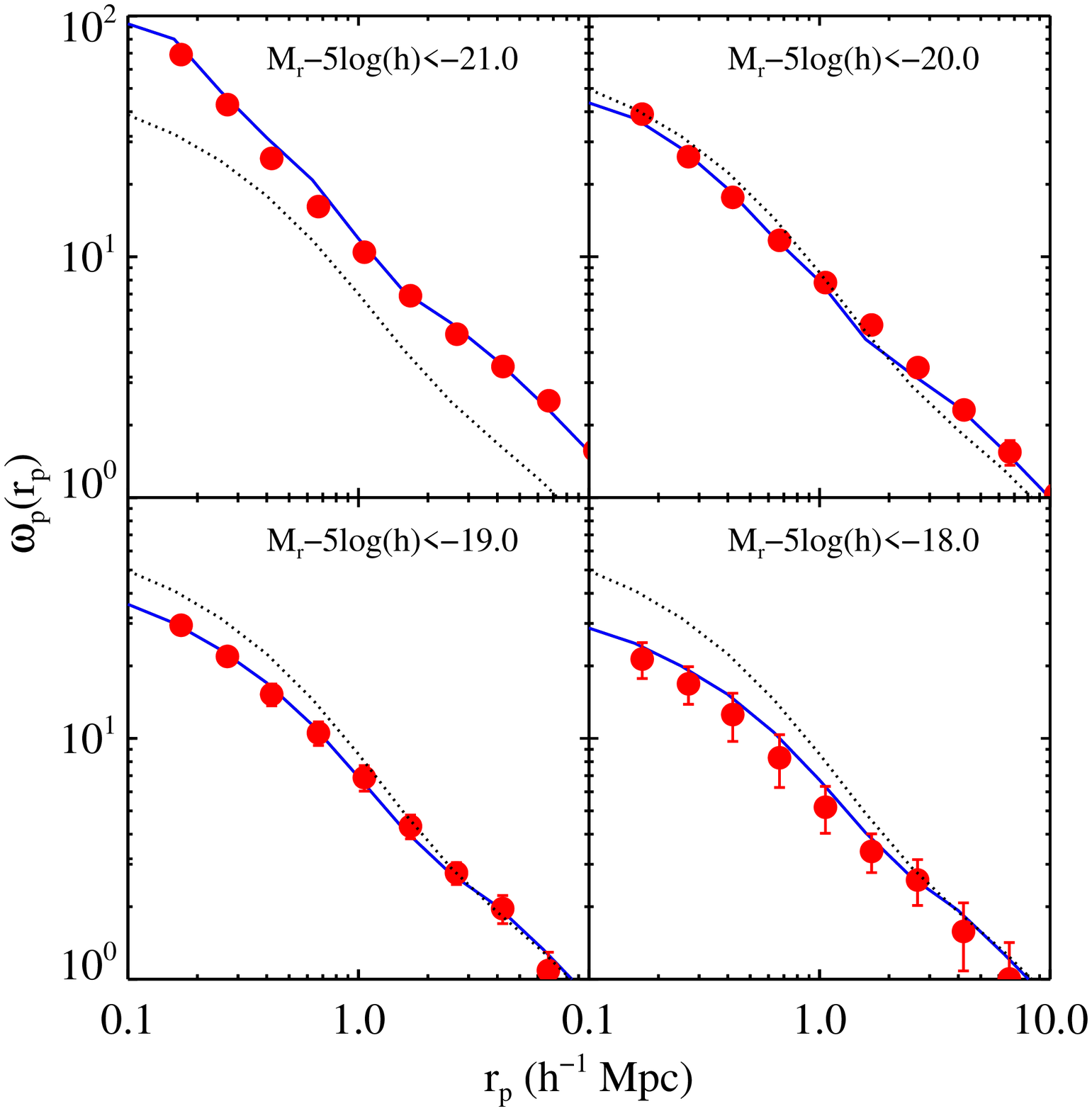,height=3in}
\psfig{figure=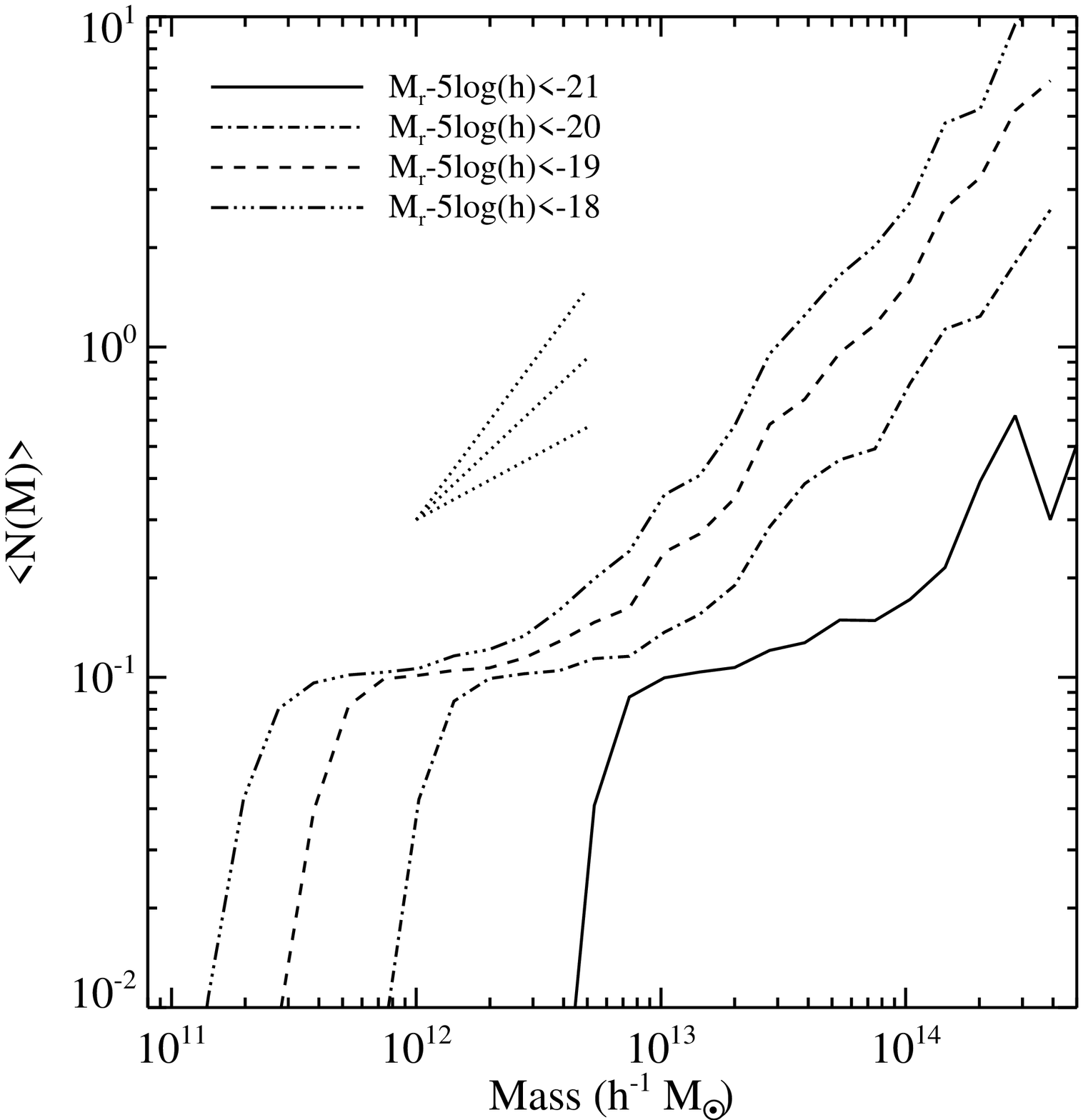,height=3in}
\caption{Left: comparison between the SDSS projected correlation
  function (points) and the correlation function derived from halos
  (solid lines) for various luminosity threshold samples. For
  comparison we include the correlation function of dark matter
  particles (dotted lines) at the median redshift of the sample. Right: the meannumber of galaxies in halos of different masses for the luminosity 
samples shown in the left panels. \hspace{4cm}  Reproduced from Conroy et al. (2006). 
\label{fig:sdss}}
\end{figure}

An important feature of the HOD shown in Figure~\ref{fig:sdss} is the
``shoulder'' near the minimum mass of the sample. The total HOD in the
region of the shoulder is narrower than the Poisson distribution ---
the fact that produces a nearly power-law correlation function of
galaxies at lower redshifts.\cite{benson_etal00,berlind_etal03} At
higher redshifts, the shortening and steepening of the shoulder due to
the younger age of host halos and more frequent presence of massive
subhalos results in strong departures from power-law correlation
function at small scales (see \S~\ref{sec:highz} and
Figure~\ref{fig:highz}).

Remarkably, the form of the HOD derived for subhalos in
dissipationless simulations is very similar to galaxy HOD measured in
$N$-body+gasdynamics SPH simulations and models employing
semi-analytic models with dissipationless
simulations.\cite{berlind_etal03,zheng_etal05,weinberg_etal06} This
supports the general framework of a close galaxy-halo connection. An
additional important implication is that the physics shaping the HOD
is relatively simple and is not sensitive to the details and specific
assumptions of galaxy formation model.

If halo circular velocity correlates with the luminosity of galaxies
they harbor, we can expect that galaxy occupation distribution
should have a similar form. Indeed, halo model fits to the galaxy clustering
measurements in the SDSS survey using 
the HOD form described above provide an excellent description of the
data.\cite{Zehavi05,abazajian_etal05} Moreover, the HOD parameters
derived from the data fits are in general agreement with the values expected
from simulations.\cite{Zehavi05}

\subsection{Modeling luminosity-dependent clustering in the SDSS survey}
\label{sec:sdss}

Galaxy clustering as a function of luminosity at $z=0$ predicted using
halo catalogs from dissipationless simulations and a simple,
non-parametric model relating galaxy luminosity to halo $V_{\rm max}$,
described in \S~\ref{sec:ghrel} is shown in the right panels of
Figure~\ref{fig:sdss}.  The figure shows remarkably good agreement
between predictions and clustering of the SDSS galaxies at all
luminosities. Note that both the normalization and shape of the correlation
function are reproduced.  It is critical to realize that the agreement
on scales $r_p\lesssim 1 h^{-1}$ Mpc is due to the luminosity
assignment scheme using $V_{\rm max}^{\rm acc}$. The luminosity
assigned using present $V_{\rm max}$ for subhalos would result in a
significant under-prediction of amplitude of $\omega_{\rm p}$ at small
scales, especially for fainter samples.

At the same time, the figure shows that galaxies of different
luminosity exhibit different amplitude and scale dependence of their
bias with respect to the overall matter correlation function in
simulation.  Although the bias appears to be complicated and scale-
and luminosity-dependent, it is faithfully reproduced by halos.

The right panel of Figure~\ref{fig:sdss} shows the first moment of the
HOD (the mean number of galaxies in halos of mass $M$) for the
luminosity samples shown in the left panels. The HOD shows explicitly
that galaxies of higher luminosity reside in more massive halos. The
overall shape of the HOD, however, is similar for all luminosities.

\subsection{Modeling galaxy clustering through cosmic time}
\label{sec:highz}

\begin{figure}
\vspace{-0.5cm}
\psfig{figure=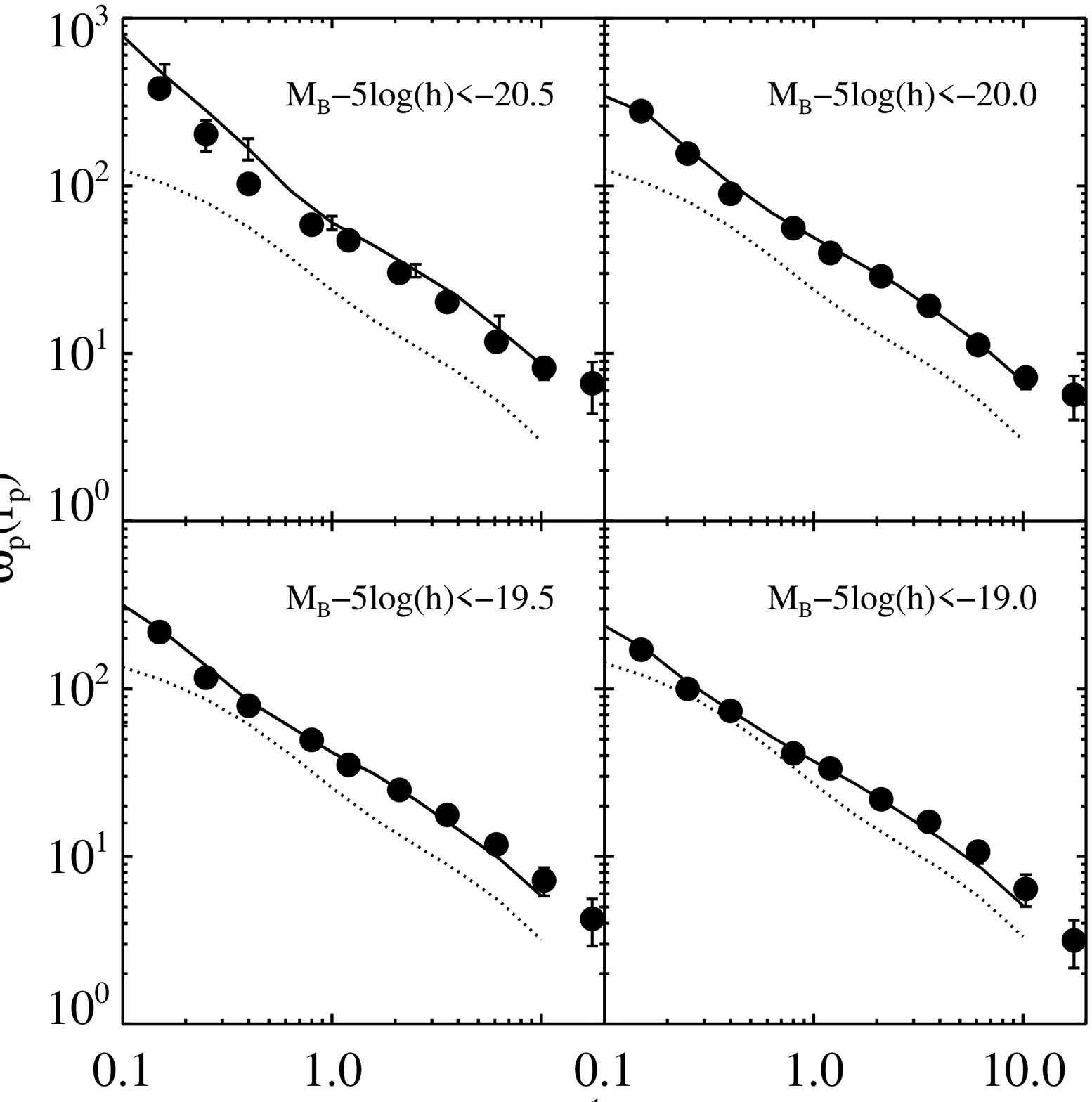,height=3in}
\psfig{figure=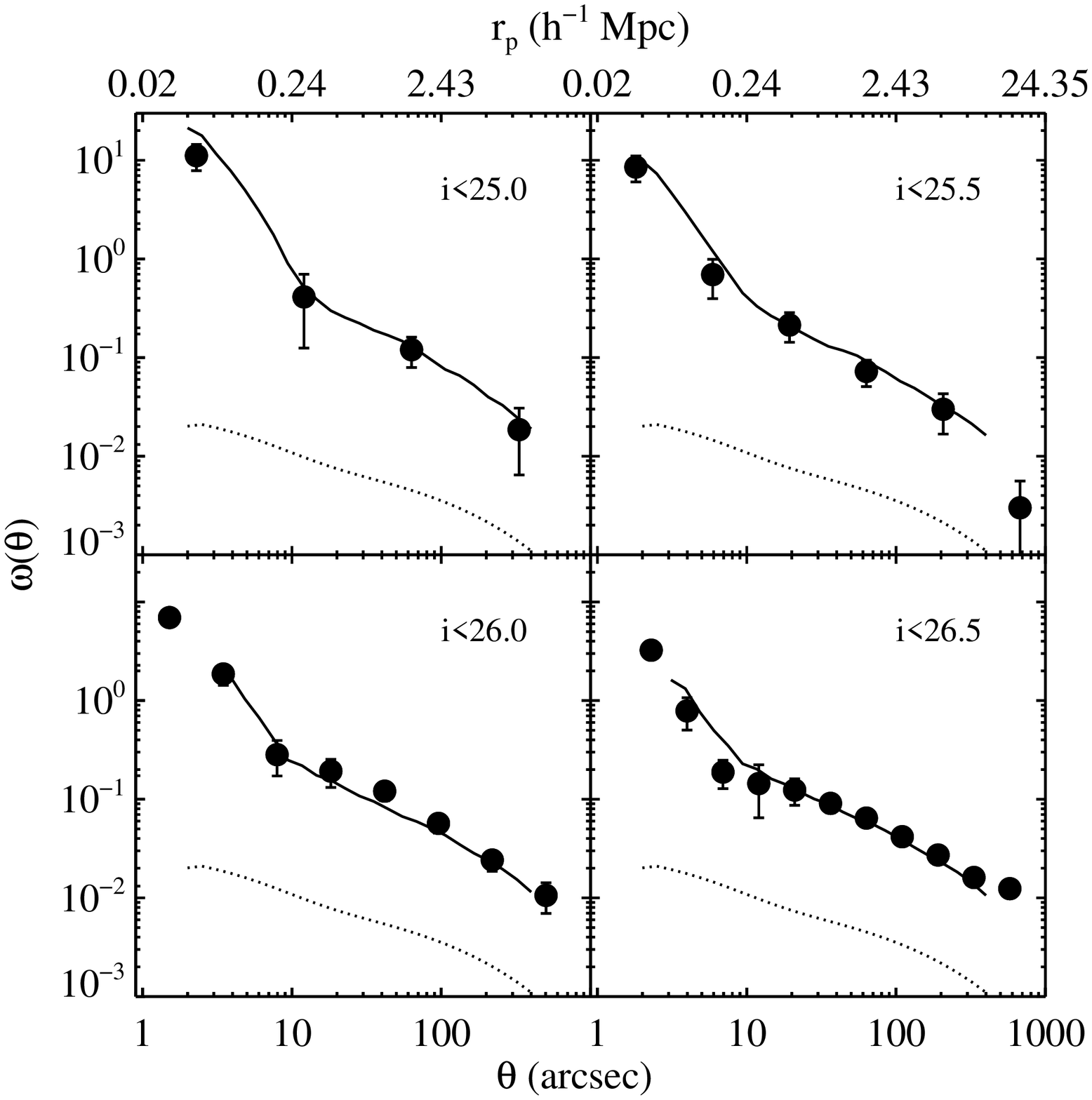,height=3in}
\caption{Left: comparison between the DEEP2 projected correlation
  function (points) and the correlation function derived from halos
  (solid lines) for various luminosity threshold samples. For
  comparison we include the correlation function of dark matter
  particles (dotted lines) at the median redshift of the sample.
  Right: similar comparison for the angular correlation function of
  the Lyman Break Galaxies with different apparent magnitude limits in
  the Subaru survey at $z\sim 4$. Note that the galaxy correlation function
strongly deviates from the power law at small scales, the behavior
expected for high-$z$ objects. The dotted lines show the corresponding
correlation function of matter in the simulation. At these high redshifts
the LBG galaxies are highly biased with respect to the matter distribution. 
 Reproduced from Conroy et al. (2006).
\label{fig:highz}}
\end{figure}

Figure~\ref{fig:highz} shows similar comparisons with galaxy
clustering as a function of luminosity at $z\sim 1$ in the DEEP2
survey\cite{Coil05} and clustering of Lyman Break Galaxies (LBGs) in
the Subaru survey\cite{Ouchi05} at $z\sim 4$. At both redshifts, the
overall agreement is again excellent on all scales.  Small
discrepancies at $r_{\rm p}\lesssim 0.5h^{-1}$~Mpc for the $M_B-5{\rm log}h<-20.5$ 
DEEP2 sample may be attributed to cosmic variance and
poisson noise.

It is worth stressing again that this remarkable agreement between
observed and model clustering is achieved using the halo distribution
in {\it dissipationless} simulations with a simple, non-parametric
relation between galaxy luminosity and halo circular velocity.  The
luminosity-dependent bias at all redshifts hence seems to be driven
entirely by the fact that brighter galaxies reside in more massive
halos, with the correspondence between halo and luminosity determined
by matching the observed luminosity function to the dark matter halo
circular velocity function.  This may appear as a reasonable and not
unexpected result for lower redshift galaxies where luminosity may be
expected to be a good tracer of stellar mass. The agreement for the
LBG galaxies is more surprising and suggests that, like galaxies in
lower redshift samples, the LBGs are fair tracers of the overall halo
population and that their restframe UV luminosity is tightly
correlated with the circular velocity (and hence mass) of their dark
matter halos.

Note that at $z\sim 4$ the correlation function of the LBG galaxies
exhibits strong departures from the power law at small scales, the
behavior that was expected based on the halo model arguments and
simulations.\cite{Zheng04,kravtsov_etal04} Similar behavior
was also recently observed in clustering of LBG galaxies in the GOODS
survey at $z=4$ and $z=5$.\cite{Lee05}

The analysis of the HOD of the $z=4$ halos in the simulation supports
the model in which most LBGs are the central galaxies in their host
halos with luminosity tightly related to the halo circular velocity
and mass.  Most LBGs have no neighbors within the same halo. However,
a fraction of them do and it is this fraction that is responsible for
the strong upturn in the correlation function at small scales. The
reason that small-scale upturn in the correlation function becomes
more pronounced at higher redshifts is that a larger fraction of the
high-$z$ halos has a satellite of comparable mass relative to the
lower-$z$ objects.  In terms of the HOD, these differences manifest in
a shorter and less flat ``shoulder'' of $\langle N(M)\rangle$ near the
minimum mass of the sample at higher redshifts.

By accurately reproducing both the small-scale upturn in
$\omega(\theta)$ and the large-scale clustering, our model accurately
predicts not only the correct distinct halos to associate with LBGs
(the `2 halo term' in halo model jargon) but also the number of LBGs
within a distinct halo (the corresponding `1 halo term').

\subsection{Are we missing galaxies in dissipationless simulations?}
\label{sec:orphans}

\begin{figure}
\vspace{-0.5cm}
\begin{center}
\psfig{figure=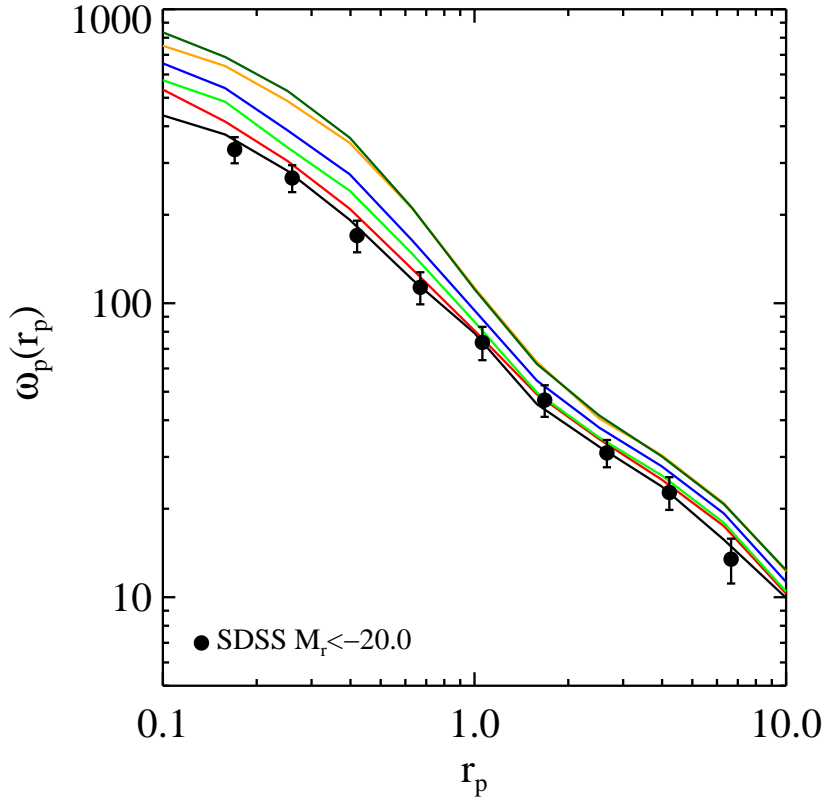,height=4.5in}
\end{center}
\vspace{-0.5cm}
\caption{Comparison of the projected two-point correlation function for
  SDSS galaxies with luminosities $M_r<-20$ and predictions from
  dissipationless simulations with different assumptions about
  fraction of orphan satellite galaxies (galaxies with subhalos
  disrupted by tides, See~\S~\ref{sec:orphans}). The solid line black line
consistent with the data is the 
prediction without orphans (i.e., using only subhalos in the simulation; 
this is the same line 
as in Figure~\ref{fig:sdss}). Other lines correspond to orphan fraction
increase by 10\% (the lines thus show effect of orphan fractions between
 10\% and 50\%). Orphan fractions of $\gtrsim 10\%$ modify the amplitude 
and shape of $w_{\rm p}$ sufficiently to break the agreement 
with the data. Figure courtesy of
  S.C.~Conroy.
\label{fig:orphans}}
\end{figure}

In dissipationless simulations, subhalos experience tidal heating and
mass loss by the tidal field of the host halo. Such tidal mass loss
and disruption are not numerical artefacts but are real physical
processes affecting evolution of
subhalos.\cite{moore_etal96,klypin_etal99,diemand_etal04} However, one
can argue that baryonic dissipation effects may greatly enhance
resistance of galaxies and their halos to tidal disruption. A decision
thus has to be made as to subsequent evolution of galaxy when its
subhalo is disrupted by host halo tides.\footnote{Operationally,
disruption should be understood as a significant tidal mass loss
which brings the total mass of a halo below a threshold of a halo
 sample or resolution of the simulation.} One extreme is to assume
that stellar components of galaxies are always resistant to tidal
disruption and are thus never disrupted. Such an assumption leads to
up to $\approx 30-40\%$ of satellite galaxies without subhalos in
dissipationless simulations.\cite{gao_etal05,wang_etal06} 
 Another extreme is to assume that stellar systems get
tidally disrupted at the same time or shortly after the host halo is disrupted
(as is assumed in the model used for comparisons with data above).

In order to assess the quantitative impact of such possible ``orphan''
galaxies on clustering statistics, the following simple test can be
performed. The fraction of subhalos in a simulation is increased in
such a way that the subhalo occupation function, $\langle
N(M)\rangle_{\rm sat}$, increases in amplitude while maintaining the
same shape.  In order to match the observed galaxy luminosity
function, the overall $V_{\rm max}$ threshold for a given sample is
simultaneously increased, such that the number density does not
change. Figure~\ref{fig:orphans} shows the predicted projected 2-point
correlation function of galaxies with $M_r<-20$ (results are similar
for other luminosities) with different assumed fractions of orphan
satellite galaxies. Even a small, $\gtrsim 10\%$, contribution of
orphan galaxies breaks the excellent agreement of the model with the
data. Note that the amplitude of the correlation function is more
sensitive to the presence of orphan galaxies at small scales ($r_{\rm
  p}\lesssim 1h^{-1}$~Mpc). Thus, addition of a significant fraction of the
orphans changes not only the amplitude, but also the shape of the
correlation function.

The existence of a significant ($\sim 30-40\%$) fraction of
orphan galaxies in the real universe would imply that 
the excellent agreement with the observed amplitude and shape of
the correlation function at different luminosities and at different
redshifts is fortuitous. The disruption of stellar systems by 
tides is therefore an important problem which should be tackled
with high-resolution numerical simulations that include stellar
component.

\subsection{Future prospects}
\label{sec:prospects}

The presented results indicate that the clustering can be modeled quite
successfully with dissipationless simulations. As the amount and
quality of data on galaxy clustering continues to improve, we can
expect more stringent model tests. It would not be surprising if data
soon reveals limitations of such a simple model. As I noted above, an
interesting avenue is to perform joint comparison of the model to the
galaxy-mass and galaxy-galaxy correlations.  Such comparison can
potentially constrain the level of scatter in the $L-V_{\rm max}$
relation.

On the theoretical side, comparisons with data can be done with
simulations of larger volumes to reduce statistical and cosmic
variance errorbars.  Smaller error bars should allow us to see subtle
deviations from the data. At the same time, the $L-V_{\rm max}$ model
used in dissipationless simulations can be directly and thoroughly
tested with $N$-body+hydro simulations with cooling and star
formation. One important question that can be addressed in
simulations is tidal disruption of galaxies and their DM halos
and existence of ``orphan'' galaxies. This work is currently underway.

\section*{Acknowledgments}
I would like to thank my collaborators Charlie Conroy, Risa Wecshler, 
Iro Tasitsiomi, Daisuke Nagai, and Andreas Berlind for fun and rewarding collaborations
which produced results described here. I would also like to thank 
Andrew Zentner and Jeremy Tinker for many stimulating discussions 
on halo model. Special ``thank you'' goes to Anatoly Klypin 
and Stefan Gottl\"ober for 
running many of the simulations used in the analyses presented here, many 
years of fruitful collaboration, and
many enlightning discussions. 
This work was supported by the National Science Foundation (NSF)
under grants No.  AST-0206216, AST-0239759 and AST-0507666, and by
NASA through grant NAG5-13274. This research was carried out at the
University of Chicago Kavli Institute for Cosmological Physics (KICP) and
was supported in part by the grant NSF PHY-0114422. KICP is an NSF
Physics Frontier Center.

\end{document}